\documentclass[10pt]{article}
\usepackage[margin=.8in,left=.8in]{geometry}

\usepackage{amsthm}
\usepackage{amssymb}
\usepackage{amsmath}
\usepackage{mathtools}
\usepackage{adjustbox}

\usepackage[table]{xcolor}

\usepackage{stmaryrd}    
\usepackage{xfrac}
\usepackage{enumitem}\setlist{nosep}

\usepackage[safe]{tipa} 

\usepackage{hhline}

\usepackage{tensor}

\usepackage{tikz}
\usetikzlibrary{
  backgrounds, 
  decorations.pathmorphing, 
  decorations.markings,
  decorations.text
}
\usetikzlibrary{cd}

\usepackage{mathrsfs} 
\DeclareMathAlphabet{\mathpzc}{OT1}{pzc}{m}{it} 

\usepackage{hyperref}
\hypersetup{
  colorlinks = true,
  linkcolor= darkblue, citecolor=darkblue
}

\hypersetup{
        colorlinks=true,
        linkcolor=darkblue,
        filecolor=darkblue,      
        urlcolor=black,
        citecolor=darkblue,
        linktoc=page
    }

\usepackage{cleveref}

\definecolor{darkblue}{rgb}{0.05,0.25,0.65}
\definecolor{darkgreen}{RGB}{20,140,10}
\definecolor{lightgray}{rgb}{0.9,0.9,0.9}
\definecolor{darkorange}{RGB}{200,100,5}
\definecolor{darkyellow}{rgb}{.91,.91,0}
\definecolor{orangeii}{RGB}{200,100,5}
\definecolor{lightblue}{RGB}{243, 250, 255}

\definecolor{lightolive}{RGB}{225, 220, 185}

\theoremstyle{definition}

\usepackage{accents}
\newlength{\dhatheight}

\let\PLAINthebibliography\thebibliography
\renewcommand\thebibliography[1]{
  \PLAINthebibliography{#1}
  \setlength{\parskip}{0.5pt}
  \setlength{\itemsep}{0.5pt plus .3ex}
}

\newcommand{\proofstep}[1]{\scalebox{.85}{#1}}

\newcommand{\yields}{\Rightarrow}

\newcommand{\ZTwo}{\mathbb{Z}_2}

\newcommand{\defneq}{\equiv}

\newcommand\bos[1]{\mathstrut\mkern2.5mu#1\mkern-14mu\raise1.7ex%
  \hbox{$\scriptstyle\rightsquigarrow$}}

\newcommand\bosonic[1]{\mathstrut\mkern2.5mu#1\mkern-14mu\raise1.7ex%
  \hbox{$\scriptstyle\rightsquigarrow$}}

\usepackage[cmtip,all]{xy}
\newcommand{\longsquiggly}{\xymatrix{{}\ar@{~>}[r]&{}}}

\usepackage{bbm} 

\usepackage[new]{old-arrows}   

\newcommand{\grayunderbrace}[2]{\color{gray}\underbrace{\color{black}#1}_{\color{gray}#2}\color{black}}

\begin{document}

\setlength{\abovedisplayskip}{3pt}
\setlength{\belowdisplayskip}{3pt}
\setlength{\abovedisplayshortskip}{-4pt}
\setlength{\belowdisplayshortskip}{3pt}




\title{The M-algebra completes the hierarchy
\\
of Super-Exceptional Tangent Spaces }

\author{
  \def\arraystretch{.9}
  \begin{tabular}{c}
  Grigorios Giotopoulos${}^{a}$,
  \\
  \footnotesize
  \tt gg2658@nyu.edu
  \end{tabular}
  \;\;
  \begin{tabular}{c}
  Hisham Sati${}^{a, b}$,
  \\
  \footnotesize
  \tt hsati@nyu.edu
  \end{tabular}
  \;\;
  \begin{tabular}{c}
  Urs Schreiber${}^{a, \dagger}$
  \\
  \footnotesize
  \tt us13@nyu.edu
  \end{tabular}
}

\maketitle

\begin{abstract}
 \noindent  The conjectured symmetries of M-theory famously involve 
   (1.) brane-extended super-symmetry (the M-algebra) and 
  (2.) exceptional duality symmetry (the $\mathfrak{e}_{11}$-algebra); 
  but little attention has been given to their inevitable combination. 
  
  \smallskip 
  In this short note, we highlight (by combining results available in the literature) that the {\it local} exceptional duality symmetry (the hyperbolic involutory $\mathfrak{k}_{1,10} \subset \mathfrak{e}_{11}$) acts on the M-algebra, through the ``brane-rotating symmetry'' $\mathfrak{sl}_{32}$, in a way which extends the known hierarchy of finite-dimensional $n$-exceptional tangent spaces compatibly beyond the traditional bound of $n \leq 7$ all the way to $n = 11$.
\end{abstract}

\vspace{.8cm}

\begin{center}
\begin{minipage}{5cm}
  \tableofcontents
\end{minipage}
\end{center}

\medskip

\vfill

\noindent
{\bf Keywords.} Kac-Moody algebras, supergravity, supersymmetry, generalized geometry, U-duality, M-theory

\vspace{.6cm}

\hrule
\vspace{5pt}

{
\footnotesize
\noindent
\def\arraystretch{1}
\tabcolsep=0pt
\begin{tabular}{ll}
${}^a$\,
&
Mathematics, Division of Science; and
\\
&
Center for Quantum and Topological Systems,
\\
&
NYUAD Research Institute,
\\
&
New York University Abu Dhabi, UAE.  
\end{tabular}
\hfill

\vspace{1mm} 
\noindent ${}^b$The Courant Institute for Mathematical Sciences, NYU, NY.

\vspace{.1cm}
\noindent ${}^\dagger$ Corresponding author

\vspace{.2cm}

\noindent
The authors acknowledge the support by {\it Tamkeen} under the 
{\it NYU Abu Dhabi Research Institute grant} {\tt CG008}.
}

\newpage

\section{Introduction}
\label{Introduction}

\noindent
{\bf Overview.}
We have previously explored 
(\cite[\S 4.5]{FSS20-HigherT}\cite{FSS20Exc}\cite{FSS21Exc}, following \cite{Vaula07}\cite{Bandos17})
the ``hidden M-algebra'' (\cite{DF82}\cite{BDIPV04}, cf. \cite{Sezgin97}\cite[\S 5]{AndrianopoliDAuria24}) as a candidate {\it super-exceptional} Kleinian model geometry (cf. \cite{Sharpe97}) for M-theory; but the relation to exceptional duality-symmetry had remained somewhat sketchy \cite{Vaula07}. 

\smallskip 
Here we observe (see \eqref{TowerOfExceptionalTangentBundles} below) 
that the familiar hierarchy of finite-dimensional exceptional tangent bundles with $\mathfrak{e}_n$-symmetry actually extends beyond the traditional bound of $n \leq 7$ all the way to $n \leq 11$ --- where it is super-symmetrized by the M-algebra --- if one understands that it is only the {\it local} symmetries which should act on local Kleinian model spaces.

\medskip 
The result follows by combining a number of observations in the (recent) literature (essentially all by the exceptional group at AEI Potsdam), which however does not seem to have been brought together in this way before.

\medskip




    





    





\noindent
{\bf The problem of formulating M-theory.}
After a burst of activity on aspects of M-theory in the 1990s (cf. \cite{NicolaiHelling98}\cite{Duff99-MTheory}) and the development of some piecemeal approaches (cf. \cite{Carlevaro06}) later researchers have lamented (e.g., \cite[p. 43]{Moore14}\cite{Duff20}) lack of  investigation into the actual formulation of the theory. 

\smallskip 
One exception is a program \cite{DHN02}\cite{DamourNicolai05} (review in \cite{KleinschmidtNicolai06-Review}\cite{Nicolai24})
aimed at understanding M-theory as spinorial quantum mechanics on the exceptional Kac-Moody group $E_{10}$ quotiented by its involutory maximal-compact subgroup $K(E_{10})$, a variant of a related program \cite{West17} for $E_{11}$ (going back to the original such suggestion of \cite{West01}, for review see \cite[\S 17]{West12}). It is progress towards bringing in the fermions into this picture (exposition in \cite{Kleinschmidt09}) that we draw from in \S\ref{OnGeometry} below.

\smallskip 
While in these ambitious approaches spacetime all but disappears into the duality group structure (with consequences that are currently understood only marginally),
the related program of {\it exceptional field theory} (\cite{HohmSamtleben13b}, cf. \cite{BossardKleinschmidtSezgin21}) aims to make manifest at least parts of this $E_{11}$-symmetry while retaining spacetime, by (drastically) extending the latter. It is a particular choice of such {\it exceptional-geometric} spacetime model that we are after here.

\smallskip
We are ultimately motivated by the  formulation of {\it flux quantization laws}
for M-theory 
(see \cite{SS24-Flux}\cite{FSS20-H}), controlling its global topological behaviour. Since admissible flux-quantization laws turn out \cite{GSS24-SuGra}\cite{GSS24-FluxOnM5} to be controlled by the Bianchi identities of duality-symmetric super-flux densities {\it on super-space}, this suggests that it is not just exceptional geometry but ``{\it super-exceptional geometry}'' or ``{\it exceptional super-geometry}'' which ultimately supports both the local as well as the global hidden structure of M-theory.

\medskip

\noindent
{\bf Manifesting hidden symmetries of M-theory.}
Among the few facts about M-theory known with some certainty is, famously \cite[\S 2.2-3]{Witten95}\cite{Duff96}, that its low-energy effective field theory is $D=11$, $\mathcal{N}=1$ super-gravity (aka 11D SuGra \cite[\S 1]{Duff99-MTheory}, reviews include \cite{MiemiecSchnakenburg06}\cite[\S 3]{GSS24-SuGra}).
Like all theories coupling fermions to gravity, 11D SuGra is most naturally formulated in ``1st order'' form as a theory of (super-)Cartan geometry, where the gravitational field is encoded by a (``moving'') co-frame field, namely by consistent identifications of each super-tangent space $T_x X$ of super-spacetime $X$ with a fixed local-model super-space $V$.

\smallskip 
At face value, the local model space of gravity in $1+d$ dimensions is Minkowski spacetime $V \defneq \mathbb{R}^{1,d}$ (with its canonical coordinate functions $(x^a)_{a=0}^{d}$), and a coframe field $e_x : T_x X^{1,d} \xrightarrow{\sim} \mathbb{R}^{1,d}$ is hence, locally, a $(1+d)$-tuple of differential 1-forms, $(e^a_x)_{a = 0}^{d}$, encoding a metric tensor $g$ on spacetime by their contraction with the Minkowski metric $\eta$ on the local model space: $g = \eta_{a b} \, e^a \otimes e^b$.

\smallskip 
From this perspective, it is suggestive to consider the unification of gravity with other fields by enlarging the local model space $V$ to incorporate these as generalized gravitational fields embodied by a  ``generalized co-frame'', 
hence by ``geometrizing internal degrees of freedom''.
Two disjoint versions of this idea are well-understood, and our motivation here is the {\it unification} of these two:

\smallskip 
\begin{itemize}[
  leftmargin=.7cm,
  topsep=2pt,
  itemsep=2pt
]
\item[\bf (i)] {\bf Super-geometric Gravity.} 

Extending the local model space $\mathbb{R}^{1,d}$ by a suitable spinorial $\mathrm{Spin}(1,d)$-representation $\mathbf{N}$, regarded in {\it super-odd degree} (recalled in \cite[\S 2]{GSS24-SuGra}, cf. \S\ref{SuperLieAlgebras}),  to {\it super-Minkowski spacetime}
\begin{equation}
  \label{SuperSpacetimeInIntroduction}
  \hspace{-4cm}
  \begin{tikzcd}[
    row sep=7pt,
    column sep=4pt
  ]
    \scalebox{.7}{
      \color{darkblue}
      \bf
      \def\arraystretch{.7}
      \begin{tabular}{c}
        Super
        \\
        tangent space
      \end{tabular}
    }
    &
    \mathbb{R}^{1,d} 
    \mathrlap{
      \;\times\; 
      \mathbb{R}^{0\vert \mathbf{N}}
      \;\defneq\;
      \mathbb{R}^{1,d\,\vert\,\mathbf{N}} 
    }
    \ar[d]
    \\
    \scalebox{.7}{
      \color{darkblue}
      \bf
      \def\arraystretch{.7}
      \begin{tabular}{c}
        Ordinary
        \\
        tangent space
      \end{tabular}
    }
    &
    \mathbb{R}^{1,d}
    \mathrlap{\,,}
  \end{tikzcd}
\end{equation}
hence enhancing spacetime $X^{1,d}$ to a super-spacetime $X^{1,d\,\vert\, \mathbf{N}}$,
makes the resulting generalized gravitational (co-frame) field subsume a spinorial field $\psi$ such that  $(e,\psi) : T_x X^{1,d\vert\mathbf{N}} \xrightarrow{\sim} \mathbb{R}^{1,d} \times \mathbf{N}_{\mathrm{odd}}$, which may be understood as the {\it gravitino} field of super-gravity. 

\smallskip 
Remarkably, in the case of 11D SuGra (at least), where $\mathbf{N} = \mathbf{32}$ (recalled in \S\ref{SpinorsIn11d}), all solutions to the equations of motion are fixed already by the restriction of the fields to the underlying ordinary (non-super) spacetimes (\cite[\S 3]{Tsimpis04}\cite[\S 3]{GSS24-M5Embedding}, a profound phenomenon known as ``rheonomy'' \cite[\S III.3.3]{CDF91}). This means that the passage from the ordinary local model $\mathbb{R}^{1,10}$ to its super-space version $\mathbb{R}^{1,10\,\vert\,\mathbf{32}}$ does not change the nature of the theory but serves to ``make manifest'' previously ``hidden'' symmetries: Namely, on super-spacetime $X^{1,d\vert \mathbf{N}}$ the local supersymmetry of SuGra (which requires work to prove over ordinary spacetime) simply becomes part of the general diffeomorphism invariance over super-spacetime, hence becomes ``geometrized''.

\vspace{1mm} 
\item[\bf (ii)] {\bf Exceptional-geometric Gravity.}

Moreover, 11D SuGra has subtly ``hidden'' duality symmetries \cite{CJLP98}\cite{deWitNicolai01}\cite{Samtleben23}, or {\it U-dualities} \cite{HullTownsend95}, whose manifestation is expected to go to the heart of understanding M-theory \cite{Nicolai99}\cite{ObersPiloine99}\cite{West01}.

Traditionally expected U-duality symmetry algebras $\mathfrak{e}_{n}$ are in the E-series of exceptional Kac-Moody Lie algebras (review in \cite{Cook07}), ordinarily recognized (only) after Kaluza-Klein reduction of 11D SuGra on $n$-torus fibers.

The idea of {\it exceptional-geometric} SuGra  \cite{HohmSamtleben13b} is to enhance the local model space $\mathbb{R}^{1,10} \simeq \mathbb{R}^{1,10-n} \times \mathbb{R}^n$ already before KK-reduction on $\mathbb{R}^n$ to an  {\it exceptional tangent space} \cite[\S 4]{Hull07}\cite[\S 2.1]{CSW14} with the $\mathbb{R}^n$ factor generalized to: 
\footnote{
  The distinction between $\mathbb{R}^n$ and its linear dual space $(\mathbb{R}^n)^\ast$ in \eqref{ExceptionalTangentSpaceDueToHull}, and in all of the following, serves to suggestively bring out the origin of these spaces via (double-)dimensional reduction, but is not otherwise consequential for the present discussion, since (the restriction of) the Lorentzian metric on all local model spaces considered here provides a {\it canonical} isomorphism $\mathbb{R}^n \,\simeq\, (\mathbb{R}^n)^\ast$. In the usual component notation, this is just the fact that Lorentz indices on elements like $v_a \,\in\, \mathbb{R}^{n}$, $v^a \in (\mathbb{R}^n)^\ast$ may be raised and lowered at will. 
}
\begin{equation}
  \label{ExceptionalTangentSpaceDueToHull}
  \hspace{-4cm}
  \begin{tikzcd}[
    row sep=10pt,
    column sep=4pt
  ]
  \scalebox{.7}{
    \color{darkblue}
    \bf
    \def\arraystretch{.9}
    \begin{tabular}{c}
      Exceptional 
      \\
      tangent space
    \end{tabular}
  }
  &
  \mathbb{R}^{n}
  \mathrlap{
  \,\times\,
  \wedge^2(\mathbb{R}^{n})^\ast
  \,\times\,
  \wedge^5(\mathbb{R}^{n})^\ast
  \,\times\,
  \wedge^6(\mathbb{R}^{n})
  }
  \ar[
    d,
    ->>
  ]
  \\
  \scalebox{.7}{
    \color{darkblue}
    \bf
    \def\arraystretch{.9}
    \begin{tabular}{c}
      Ordinary
      \\
      tangent space
    \end{tabular}
  }
  &
  \mathbb{R}^n
  \end{tikzcd}
  \hspace{1.4cm}
\scalebox{0.7}{$ n \in \{1, \cdots, 7\} $}
\end{equation}
on which these hidden symmetries are compatibly manifested as $\mathfrak{e}_{n}$-representations \eqref{TowerOfExceptionalTangentBundles}. Successful exceptional-geometric (re-)formulations of 11D SuGra based on the local model \eqref{ExceptionalTangentSpaceDueToHull}, and hence with manifest U-duality symmetry even without or before KK-reduction, have been constructed in \cite{HohmSamtleben13b}\cite{HohmSamtleben14a}\cite{HohmSamtleben14b} (see \cite{BermanBlair20} for a review and for the remaining cases). 

\smallskip

Previously this was mostly considered for $n \leq 7$, 
since already for $n = 8$ the dimension of \eqref{ExceptionalTangentSpaceDueToHull} is too small to support a non-trivial $\mathfrak{e}_{n}$-representations (cf. the troubles encountered for $n = 8$ in \cite[(4.4)]{HohmSamtleben14c})
and dramatically too small for $n \geq 9$, where all non-trivial $\mathfrak{e}_{n}$-representations are infinite-dimensional. We offer a resolution of this issue in \S \ref{OnGeometry}.
\end{itemize}

\smallskip

\noindent
{\bf The need for super-exceptional geometric gravity.}
While both super-geometry and exceptional-geometry are, by all indications, key to understanding M-theory, their combination to a formulation of super-exceptional gravity has received little attention (exceptions are \cite{BH79}\cite{ButterSamtlebenSezgin19}, for $n = 7$). However, in view of the somewhat miraculous emergence of (on-shell) 11D supergravity from super-geometry \cite{BH80}\cite{CF80}\cite{CDF91}\cite[Thm. 3.1]{GSS24-SuGra}, as well as indications towards the M-theoretic completion of 11D supergravity via exceptional geometry, we find it natural to look for their unification in a {\it super-exceptional} geometric formulation, involving, equivalently:
\begin{itemize}[
  leftmargin=1cm,
  topsep=1pt,
  itemsep=2pt
]
\item[\bf (i)] an exceptional-geometric enhancement of the super tangent space \eqref{SuperSpacetimeInIntroduction},

\item[\bf (ii)] a
super-geometric enhancement of the exceptional tangent space \eqref{ExceptionalTangentSpaceDueToHull}.
\end{itemize}

\medskip 
\noindent
{\bf Terminology.}
We admit to the following abuse of notation/terminology, made throughout, for convenience:

\begin{itemize}[
  itemsep=2pt,
  topsep=2pt,
  leftmargin=.4cm
]
  \item {\bf Split real forms.}
    All Lie algebras and Lie groups we refer to are {\it split real forms}. 

    By this default, we write
    \begin{itemize}[
      topsep=1pt,
      itemsep=2pt,
      leftmargin=.65cm
    ]
    
    \item[--] $\mathfrak{e}_n$ for $\mathfrak{e}_{n(n)}$ (as usual in our context, e.g. \cite{KleinschmidtNicolai06-Review}),

    \item[--] $\mathfrak{k}_n
    \,\subset\, \mathfrak{e}_n$ for its involutory (``maximal compact'') sub-algebra (as e.g. in \cite{KKLN22}\cite{LautenbacherKoehl24}), 

    \item[--] $\mathfrak{sl}_{32} 
    \,:=\, \mathfrak{sl}_{32}(\mathbb{R})\,$

    \item[--] $\mathrm{SL}(32) \,:=\, \mathrm{SL}(32; \mathbb{R})$.

    \end{itemize}
    
  \item {\bf U-duality/Hidden symmetry.}
  Even though the term ``U-duality'' refers, strictly speaking, to certain integral subgroups (denoted e.g. $E_{7}(\mathbb{Z}) \subset E_{7(7)}$ \cite{HullTownsend95}) of the duality-symmetry groups of supergravity,
  we say ``U-duality'' also for the latter (as is not unusual, cf. \cite{HohmSamtleben13a}).
\end{itemize}

\medskip

\noindent
{\bf Acknowledgements.} We are indebted to Axel Kleinschmidt and Benedikt K{\"o}nig for discussion and pointers to their work.

\section{The observation}
\label{OnGeometry}

In \S\ref{ExceptionalTangentSpaces}, we discuss how the hierarchy of finite-dimensional exceptional tangent spaces continues beyond $n\leq 7$ to $n \leq 11$,
and in \S\ref{SuperTangentSpaces} we conclude that the resulting exceptional tangent space for $n = 11$ is super-symmetrized by the M-algebra.

\subsection{Exceptional tangent spaces fully extended}
\label{ExceptionalTangentSpaces}

\noindent
{\bf Maximal compact hidden symmetry.}
To do so, we highlight the following two observations from the work of Nicolai, Kleinschmidt, et al., concerning (i) local and (ii) spinorial hidden symmetry:

\begin{itemize}[
  leftmargin=.8cm,
  topsep=1pt,
  itemsep=2pt,
]
\item[\bf (i)]
{\bf Local hidden symmetry:}
While the Kac-Moody Lie algebras $\mathfrak{e}_{n}$ reflect the expected {\it global} hidden symmetry, it is only their ``maximal compact'' (or ``involutory'') subalgebras \cite[\S 2]{KKLN22}\cite[\S 2.9]{LautenbacherKoehl24}, which reflect the corresponding {\it local} hidden symmetry:
\begin{equation}
  \label{TheMaximalCompactSubalgebras}
  \begin{tikzcd}[
    row sep=-3pt,
    column sep=20pt
  ]
  \mathllap{
  \scalebox{.7}{
    \color{gray}
    \bf
    \def\arraystretch{.9}
    \begin{tabular}{c}
      Global
      \\
      symmetry
    \end{tabular}
  }
  }
  \mathfrak{e}_{n}
\quad 
  &&
  \quad 
  \mathfrak{k}_{n}
  \ar[
    ll,
    hook'
  ]
  \mathrlap{
  \scalebox{.7}{
    \color{gray}
    \bf
    \def\arraystretch{.9}
    \begin{tabular}{c}
      Local
      \\
      symmetry
    \end{tabular}
  }  
  }
  \\
  \scalebox{.7}{
    \color{darkblue}
    \bf
    \def\arraystretch{.9}
    \begin{tabular}{c}
      Kac-Moody
      \\
      Lie algebra
    \end{tabular}
  }
  &&
  \scalebox{.7}{
    \color{darkblue}
    \bf
    \def\arraystretch{.9}
    \begin{tabular}{c}
      Maximal compact
      \\
      sub-algebra
    \end{tabular}
  }  
  \end{tikzcd}
\end{equation}

For $n=7$ and $n = 8$, this distinction is seen in the original constructions of
SuGra Lagrangians with local $\mathfrak{k}_7 \simeq \mathfrak{su}_8$-symmetry 
\cite{CremmerJulia79}\cite{deWitNicolai86}
and local $\mathfrak{k}_8 \,\simeq\, \mathfrak{so}_{16}$ symmetry \cite[p. 6]{Cremmer81}\cite{Nicolai87}.
For $n = 9$, the notion of the maximal compact $\mathfrak{k}_{9}$ playing the role of the local symmetry is highlighted in \cite[p. 7]{NicolaiSamtleben05}\cite[p. 4]{KleinschmidtNicolaiPalmkvist07}.
The general statement for all $n$ is made explicit in \cite[p. 4]{KleinschmidtNicolai21}.  \footnote{
\cite[p. 4]{KleinschmidtNicolai21}: ``Focusing on a single maximally supersymmetric theory breaks the
$E_n$ symmetry but we expect the $K(E_n)$ symmetry to remain intact, much in the same way as
the reformulations of $D = 11$ supergravity in \cite{deWitNicolai86}\cite{Nicolai87} maintain a larger local symmetry.''}, cf. also \cite[\S 3.2, p. 13]{MSCZ21}.
In fact, the emphasis on exceptional {\it local} symmetries is already visible in \cite[p.  150]{Nicolai99}.

\item[\bf (ii)] 
{\bf Spinorial hidden symmetry:}
In contrast to the Kac-Moody algebras $\mathfrak{e}_{n}$ themselves, their maximal compact subalgebras $\mathfrak{k}_{n}$ \eqref{TheMaximalCompactSubalgebras} --- while themselves still infinite-dimensional for $n \geq 9$ --- have non-trivial {\it finite-dimensional} representations \cite{KNV20}\cite{KKLN22}.

Among these is notably a 32-dimensional spinorial irrep 
both for $\mathfrak{k}_{10}$ \cite[\S 8]{dBuylHenneauxPaulot05}\cite{KleinschmidtNicolai06}\cite[\S 2.2]{KleinschmidtNicolai13} (exposition in \cite{Kleinschmidt09}) as well as for $\mathfrak{k}_{1,10}$ \cite[p 42]{BossardKleinschmidtSezgin19},
which lifts the familiar Majorana spinor representation of 11D SuGra \eqref{The11dMajoranaRepresentation}:
\begin{equation}
  \label{The32OfE10}
  \begin{tikzcd}[
    sep=0pt
  ]
    \mathfrak{so}_{1,10}
    \ar[rr, hook]
    &&
    \mathfrak{k}_{1,10}
    &&
    \mathfrak{k}_{10}
    \ar[
      ll, 
      hook'
    ]
    \\
    \mathbf{32}
    &\longmapsfrom&
    \mathbf{32}
    &\longmapsto&
    \mathbf{32}
    \,.
  \end{tikzcd}
\end{equation}

\end{itemize}

\smallskip

\noindent

Noting that the symmetry of the usual local model spaces \eqref{SuperSpacetimeInIntroduction} of 11D super-gravity is $\mathrm{Spin}(1,10)$ and hence certainly the corresponding {\it local and spinorial} symmetry,
these two observations suggest that it may be misguided to generally ask the exceptional tangent spaces like \eqref{ExceptionalTangentSpaceDueToHull} to be acted on by $\mathfrak{e}_{n}$ (as traditionally assumed) instead of by $\mathfrak{k}_{n}$ \eqref{TheMaximalCompactSubalgebras}.

\medskip

\noindent
{\bf The boundary case $n=8$.}
As a quick plausibility check, we immediately see that this perspective resolves the apparent dimensional mismatch of \eqref{ExceptionalTangentSpaceDueToHull} at $n = 8$: Here we have
$$
  \mathrm{dim}\Big(
    \mathbb{R}^8 
    \oplus 
    \wedge^2(\mathbb{R}^8)^\ast
    \oplus
    \wedge^5(\mathbb{R}^8)^\ast
    \oplus
    \wedge^6 \mathbb{R}^8
  \Big)
  \;=\;
  {8 \choose 1}
  \,+\,
  {8 \choose 2}
  \,+\,
  {8 \choose 5}
  \,+\,
  {8 \choose 6}
  \,=\
  120,
$$
which falls short of the dimension 248 of the basic rep of $\mathfrak{e}_{8}$ but is exactly the dimension of the basic representation of its maximal compact sub-algebra $ \mathfrak{so}_{16}$: 
\footnote{
  This phenomenon, around
  \eqref{BranchingOfThe248OfE8},
  was of course observed long ago, e.g. \cite[p. 3]{deWitNicolai01}, but there with $\mathfrak{so}_{16}$ thought of as the automorphism group of the brane-extended supersymmetry algebra, which happens to coincide with the maximal compact subalgebra for $n = 8$ but not beyond. Our point here is that it is instead the perspective of the maximal compact subalgebras that fixes the previously broken-looking pattern of the exceptional tangent spaces.
}

\vspace{-.4cm}
\begin{equation}
  \label{BranchingOfThe248OfE8}
  \begin{tikzcd}[sep=0pt]
    \mathfrak{so}_{16}
    \,\simeq\,
    \mathfrak{k}_{8}
    \ar[
      rr,
      "{ \iota }"
    ]
    &&
    \mathfrak{e}_{8}
    \\[-2pt]
    \mathbf{120}
    \oplus
    \mathbf{128}
    &
    \overset{
      \iota^\ast
    }{\longmapsfrom}
    &
    \mathbf{248}
    \mathrlap{\,.}
  \end{tikzcd}
\end{equation}

\noindent
{\bf The full exceptional tangent space.}
Therefore, this perspective leads us to ask whether the bosonic body of the {\it M-algebra} (\S\ref{SuperTangentSpaces}), which is
\begin{equation}
  \label{ExceptionalTangentBundleInIntroduction}
  \begin{tikzcd}[
    column sep=2pt,
    row sep=2pt
  ]
  \mathbf{32} 
    \underset{
      \mathclap{
      \mathrm{sym}
      }
    }{\otimes}
  \mathbf{32}
  &\simeq&
  \mathbf{11}
  &\oplus&
  \mathbf{55}
  &\oplus&
  \mathbf{462}
  &\in&
  \mathrm{Rep}_{\mathbb{R}}(
    \mathfrak{so}_{1,10}
  )
  \ar[
    d,
    "{  }"
  ]
  \\
  &\simeq&
  \mathbb{R}^{1,10}
  &\oplus&
  \wedge^2 
  (\mathbb{R}^{1,10})^\ast
  &\oplus&
  \wedge^5
  (\mathbb{R}^{1,10})^\ast
  &\in&
  \mathrm{Vect}_{\mathbb{R}}
  \mathrlap{\,,}
  \\
  \end{tikzcd}
\end{equation}
may provide exceptional-geometric tangent spaces for all $n \leq 11$, in the sense of carrying natural basic representations of the {\it local} hidden symmetry groups \eqref{TheMaximalCompactSubalgebras}. We find  now that, yes, this yields perfect agreement.

\medskip
First, by Hodge-dualizing the temporal components (cf. \cite[(2.12)]{Hull98} following Townsend) of would-be brane charges in \eqref{ExceptionalTangentBundleInIntroduction}, it is seen to be of essentially the form \eqref{ExceptionalTangentSpaceDueToHull}, 
except for one extra summand of 
$\mathbb{R}^{1,0}$ (the time-axis) and one extra summand of $\wedge^n \mathbb{R}^{10}$ (``9-brane charge'' \cite[p. 9]{Hull98}):
\begin{equation}
  \label{HodgeDualizedMAlgebra}
 \mathbf{32} 
    \underset{
      \mathclap{
      \mathrm{sym}
      }
    }{\otimes}
  \mathbf{32}
  \;\;\simeq\;\;
  \mathbb{R}^{1,0}
  \oplus
  \Big(
    \mathbb{R}^{10}
    \oplus
    \wedge^2(\mathbb{R}^{10})^\ast
    \oplus
    \wedge^5(\mathbb{R}^{10})^\ast
    \oplus
    \wedge^6\mathbb{R}^{10}
    \oplus
    \wedge^9\mathbb{R}^{10}
  \Big)
  \,.
\end{equation}
Hence, restricting the term in parenthesis along $\mathbb{R}^n \hookrightarrow \mathbb{R}^{10}$ for $n \in \{4,5,6,7\}$, this is just the typical fiber of the exceptional tangent bundles of \cite[\S 4]{Hull07} as commonly considered these days \cite[(4.4)]{BermanBlair20}.

\medskip 
However, we now see that for $n \geq 8$ the corresponding restriction of \eqref{HodgeDualizedMAlgebra} carries the basic representation of the {\it local} (maximal compact) hidden symmetry \eqref{TheMaximalCompactSubalgebras}
-- the top \& left part of the following \hyperlink{Table1}{Table 1} is essentially given in \cite[(2)]{deWitNicolai01}, the shaded bottom right corner expresses the new observation advertised here:
\begin{equation}
  \hypertarget{Table1}{} \label{TowerOfExceptionalTangentBundles}
  \adjustbox{
    raise=-2.5cm
  }{
\includegraphics[width=15cm]{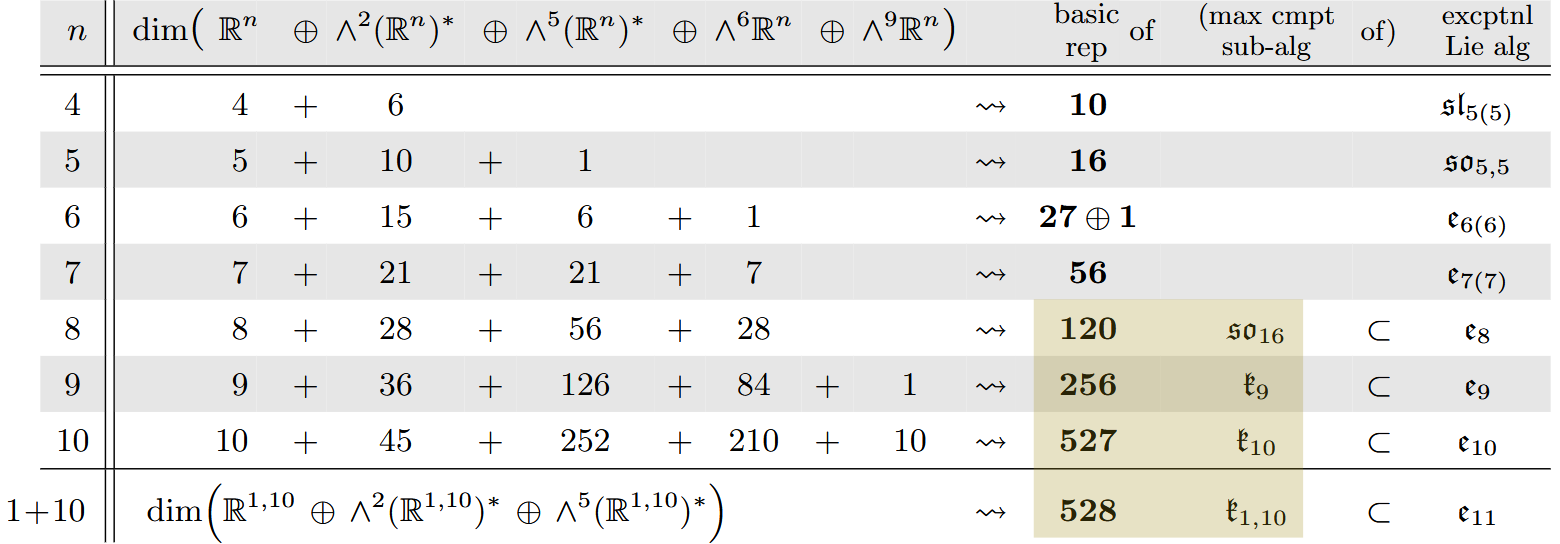}  
  }
\end{equation}
\begin{center}
  \footnotesize{
    {\bf Table 1.}
    The completed hierarchy of super-exceptional tangent spaces.
  }
\end{center}

\smallskip

The identification of the (dimensions) of basic representations in \eqref{TowerOfExceptionalTangentBundles} is given by the following facts, which is, in this combination, our main observation here:

\vspace{3mm} 
\begin{itemize}[
   itemsep=6pt,
   leftmargin=.5cm
]

\item \fbox{$\mathbf{n=4,5,6,7}$}\,: classical, e.g. \cite[\S 2]{deWitNicolai01}\cite[\S 4]{Hull07}\cite[\S 2.2]{PachecoWaldram08}\cite[\S 2.1]{CSW14};

\item \colorbox{lightolive}{$\mathbf{n = 8}$}\,: the  $\mathbf{248}$ of $\mathfrak{e}_{8}$ branches as $\mathbf{120} \oplus \mathbf{128}$ of the maximal compact $\mathfrak{so}_{16}$ ---
as a representation-theoretic statement this is classical (e.g. \cite[p. 4]{HohmSamtleben14c}),
but as part of a change in pattern from $\mathfrak{e}_{n}$ to $\mathfrak{k}_{n}$ this may not have been appreciated 
(\cite[p. 3]{deWitNicolai01} instead sees it as a partial change of pattern to the automorphism algebra of the extended susy algebra);

\end{itemize}

\smallskip

\noindent
and then the following more novel  facts (largely due to the exceptional group at the AEI in Potsdam): 
\begin{itemize}[
  topsep=6pt,
  itemsep=5pt,
  leftmargin=.5cm
]
\item  \colorbox{lightolive}{$\mathbf{n = 9}$}\,: 
the (infinite-dimensional) basic rep of $\mathfrak{e}_{9}$ branches as $\mathbf{256} \oplus \mbox{higher-parabolic-levels}$ under $\mathfrak{k}_{9}$ --- this is only very recently discussed in \cite[p. 38, 41, 42]{Koenig24};

\item  \colorbox{lightolive}{$\mathbf{n = 10}$}\,:
remarkably, there is an irrep $\mathbf{527}$ of $\mathfrak{k}_{10}$, and it appears in the symmetric square of a spinorial $\mathbf{32}$ irrep \eqref{The32OfE10}
as: 
$\mathbf{32} \otimes_{\mathrm{sym}} \mathbf{32} \,\simeq\, \mathbf{1} \oplus \mathbf{527}$
\cite[p. 37]{DKN06}, which exactly matches the interpretation here, where the bosonic dimension of the M-algebra is the same expression $\mathrm{dim}(\mathbf{32}\otimes_{\mathrm{sym}}\mathbf{32})$ --- the remaining $\mathbf{1}$ is the first summand (the time axis) in \eqref{HodgeDualizedMAlgebra};

\item  \colorbox{lightolive}{$\mathbf{n = 1+10}$}\,:
finally, re-including this temporal component and hence going back to the unbroken bosonic 

\vspace{1mm} 
\noindent M-algebra \eqref{ExceptionalTangentBundleInIntroduction}
we need an irrep $\mathbf{528}$ of $\mathfrak{k}_{1,10}$ (the hyperbolic form \cite[\S 2.2]{EnglertHouart04}
of the involutory subalgebra, denoted $H_{11}$ in \cite{Keurentjes04});
remarkably, this also exists \cite[p. 29]{GomisKleinschmidtPalmkvist19} 
and it is isomorphic to the symmetric square $\mathbf{32} \otimes_{\mathrm{sym}} \mathbf{32} \,\simeq\, \mathbf{528}$
(cf. \cite[\S D]{BossardKleinschmidtSezgin19}) of the original $\mathbf{32}$ \eqref{The32OfE10} that we started with.

\vspace{1mm} 
\noindent (A suggestion reminiscent of this action of $\mathfrak{k}_{1,10}$ on the basic M-algebra was previously made in \cite[p. 14]{Vaula07}.)
\end{itemize}

\vspace{.7cm}

In summary this suggests that the  $\mathbf{528}$ of $\mathfrak{k}_{1,10}$ is the root of the hierarchy of exceptional tangent spaces 
\eqref{TowerOfExceptionalTangentBundles}, while at the same time exactly unifying $\mathbf{11}$-dimensional spacetime with the $\mathbf{55}$ M2- and $\mathbf{462}$ M5-brane charges:
\begin{equation}
  \label{RootDiagram}
  \begin{tikzcd}[
    column sep=8pt,
    row sep=7pt
  ]
    &[+5pt]&[+5pt]
    \mathfrak{k}_{1,10}
    &[+5pt]&[+5pt]
    \\
    &&
    \mathbf{528}
    \ar[
      dd,
      <-,
      "{\sim}"{sloped}
    ]
    \\[-9pt]
    \mathclap{
      \scalebox{.7}{
        \color{gray}
        \bf
        \def\arraystretch{.9}
        \begin{tabular}{c}
          local
          \\
          Lorentz
          \\
          symmetry
        \end{tabular}
      }
    }
    && && &&
    \mathclap{
      \scalebox{.7}{
        \color{gray}
        \bf
        \def\arraystretch{.9}
        \begin{tabular}{c}
          local hidden symmetries
        \end{tabular}
      }
    }
    \\[-9pt]
    &&
    \mathbf{32}
    \underset{
      \mathclap{
        \scalebox{.6}{$\mathrm{sym}$}
      }
    }{\otimes}
    \mathbf{32}
    \\[-20pt]
    \mathfrak{so}_{1,10}
    \ar[
      uuuurr,
      hook
    ]
    && &&
    \mathfrak{k}_{10}
    \ar[
      uuuull,
      hook'
    ]
    &&
    \mathfrak{k}_{9}
    \ar[
      ll, 
      hook'
    ]
    &&
    \mathfrak{so}_{16}
    \ar[
      ll, 
      hook'
    ]
    &\phantom{--}&
    {}
    \ar[
      ll, 
      dotted,
      hook'
    ]
    \\
    \mathbf{11}
    \oplus
    \mathbf{55}
    \oplus
    \mathbf{462}
    \ar[
      uurr,
      <-,
      shorten >=-8pt,
      "\sim"{sloped}
    ]
    && &&
    \mathbf{527}
    \oplus
    \mathbf{1}
    \ar[
     uull,
     <-,
      shorten >=-8pt,
     "{ \sim }"{sloped}
    ]
    &\longmapsto&
    \mathbf{256} \oplus \cdots
    &\longmapsto&
    \mathbf{120} \oplus \cdots
    \\[-10pt]
    \mathclap{
      \scalebox{.7}{
        \def\arraystretch{.9}
        \begin{tabular}{c}
          \color{darkblue}
          \bf
          M-algebra
          \\
          (bosonic body) 
        \end{tabular}
      }
    }
    &&
    &&
    \mathrlap{
      \scalebox{.7}{
        \color{darkblue}
        \bf
        \def\arraystretch{.9}
        \begin{tabular}{c}
          exceptional
          tangent spaces
        \end{tabular}
      }
    }
  \end{tikzcd}
\end{equation}

\medskip

\noindent
{\bf Effective $\mathfrak{k}_{1,10}$-action through $\mathfrak{sl}_{32}$.} Of course, the infinite-dimensional $\mathfrak{k}_{1,10}$ acts on the finite-dimensional $\mathbf{528}$
\eqref{RootDiagram} through a finite-dimensional quotient Lie algebra, which turns out to be $\mathfrak{sl}_{32}$ \cite[p. 42]{BossardKleinschmidtSezgin19}.
This had previously been noticed as an automorphism symmetry of the basic M-algebra (we recall this below on p. \pageref{ManifestGL32}) and called the {\it brane-rotating symmetry} \cite{BaerwaldWest00} because it mixes (spacetime with) the M-brane components (cf. Ex. \ref{BraneRotatingSymmetry} below).

\medskip

To make this fully explicit, we recall now the M-algebra as a super-space enhancement of the $n=11$-exceptional tangent space  $\mathbb{R}^{1,10}\,\oplus\, \wedge^2(\mathbb{R}^{1,10})^\ast \,\oplus\, \wedge^5 (\mathbb{R}^{1,10})^\ast$ \eqref{ExceptionalTangentBundleInIntroduction}, and how it inherits its $\mathfrak{sl}_{32}$-action and hence its $\mathfrak{k}_{1,10}$-action.

\medskip

\subsection{Super-tangent space fully extended}
\label{SuperTangentSpaces}

\noindent
{\bf The super-Minkowski algebra.}
By the  ($D=11$, $\mathcal{N}=1$) {\it super-Minkowski Lie algebra} we mean the super-translational super-Lie sub-algebra of the super-Poincar{\'e} algebra 
\footnote{
The full super-Poincar{\'e} super Lie algebra (aka: ``supersymmetry algebra'') is the semi-direct product $\mathbb{R}^{1,10} \rtimes \mathfrak{so}(1,10)$ of the super-Minkowski algebra \eqref{SuperMinkowskiLinearBasis} with the Lorentz Lie algebra $\mathfrak{so}(1,10)$ acting on $\mathbb{R}\big\langle (P_a)_{a = 0}^{10}\big\rangle$ as its defining/vector representation and on $\mathbb{R}\big\langle (Q_\alpha)_{\alpha=1}^{32} \big\rangle \,\simeq\, \mathbf{32}$ as its irreducible Majorana spin representation \eqref{The11dMajoranaRepresentation}. Similarly, there is the semidirect product with $\mathfrak{so}(1,10)$ of the basic M-algebra \eqref{TheBasicMAlgebra}, which may be regarded as the full M-symmetry algebra, see \hyperlink{TableKleinian}{Table 1}. But since no further subtleties are involved in forming these semidirect products with the Lorentz algebra, we do not further dwell on them here.} 
(commonly known as the {\it supersymmetry algebra}) whose underlying super-vector space is (cf. our super-algebra conventions in \S\ref{SuperLieAlgebras})
\begin{equation}
  \label{SuperMinkowskiLinearBasis}
  \mathbb{R}^{1,10\,\vert\,\mathbf{32}}
  \;\simeq\;
  \mathbb{R}\Big\langle \!
     \grayunderbrace{
       (Q_\alpha)_{\alpha=1}^{32}
     }{ 
       \mathrm{deg}
       \,=\, 
       (0,\mathrm{odd}) 
    }
    \,,
    \grayunderbrace{
      (P_a)_{a = 0}^{10}
    }{ 
      \mathrm{deg} \,=\, (0,\mathrm{evn}) 
    }
  \!\!\Big\rangle
\end{equation}
with the only non-trivial super-Lie brackets on basis elements being \footnote{
  \label{PrefactorConvention}
  Our prefactor convention in \eqref{TheBifermionicSuperBracket} -- ultimately enforced via the translation 
\eqref{RelationBetweenStructureConstants}
  by our convention for the super-torsion tensor in \cite{GSS24-SuGra}\cite{GSS24-FluxOnM5} and \cite{GSS24-M5Embedding} --
  coincides with that in \cite[(1.16)]{DeligneFreed99}\cite[p. 52]{Freed99}.
}
\begin{equation}
  \label{TheBifermionicSuperBracket}
  \big[
    Q_\alpha
    ,\,
    Q_\beta
  \big]
  \;=\;
  -
  2\,
  \Gamma^a_{\alpha \beta}
  P_a
  \,.
\end{equation}

\smallskip

Its Chevalley-Eilenberg algebra \eqref{RelationBetweenStructureConstants}
therefore has the underlying graded super-algebra
\vspace{1mm}
\begin{equation}
  \label{CEOfSuperMinkowksi}
  \mathrm{CE}\big(
    \mathbb{R}^{1,10\,\vert\,\mathbf{32}}
  \big)
  \;\simeq\;
  \mathbb{R}\big[\!
    \grayunderbrace{
     (\psi^\alpha)_{\alpha = 0}^{32}
    }{
      \mathrm{deg}\,=\, (1,\mathrm{odd})
    }
    ,\,
    \grayunderbrace{
    (e^a)_{a=0}^{10}
    }{ 
      \mathrm{deg} = (1,\mathrm{evn})
    }
  \!\big]
\end{equation}
with the differential given on generators by

\vspace{-.4cm}
\begin{equation}
  \label{CEDifferentialForSuperMinkowski}
  \begin{array}{cll}
    \mathrm{d}\, \psi
    &=&
    0
    \\
    \mathrm{d}\, 
    e^a
    &=&
    \big(\hspace{1pt}
      \overline{\psi}
      \,\Gamma^a\,
      \psi
    \big)
    \,.
  \end{array}
\end{equation}

\vspace{1mm} 
For the following, it is instructive to note that the 2-forms $\big(\hspace{1pt}\overline{\psi}\,\Gamma^a\,\psi\big) \in \mathrm{CE}(\mathbb{R}^{0\,\vert\,\mathbf{32}})$ are non-trivial 2-cocycles on the purely fermionic abelian subalgebra $\mathbb{R}^{0\vert \mathbf{32}}$ --- the {\it super-point} -- whence \eqref{CEDifferentialForSuperMinkowski} exhibits the super-Minkowski algebra as a central extension of the superpoint (cf. \cite[\S 2.1]{CdAIPB00}\cite{HS18}):
\begin{equation}
  \label{SuperMinkowskiAlgebrasAsCentralExtension}
  \begin{tikzcd}[
    sep=10pt
  ]
    0
    \ar[r]
    &
    \mathbb{R}^{1,10}
    \ar[rr, hook]
    &&
    \mathbb{R}^{1,10\,\vert\,\mathbf{32}}
    \ar[rr, ->>]
    &&
    \mathbb{R}^{0\,\vert\,\mathbf{32}}
    \ar[r]
    &
    0
    \,.
  \end{tikzcd}
\end{equation}

\smallskip

\noindent
{\bf The basic M-algebra.}
Concerning $\big(\hspace{1pt}\overline{\psi}\,\Gamma^a\,\psi\big)$ in \eqref{CEDifferentialForSuperMinkowski} being a 2-cocycle, it is obvious that it is closed  and not exact --- since $\psi$ is closed and not exact \eqref{CEDifferentialForSuperMinkowski} --- but what is mildly non-trivial is that it exists as a non-vanishing $\mathrm{Spin}(1,10)$-invariant 2-form in the first place: The only further expressions for which this is the case are 
\begin{equation}
  \label{Further2CocyclesOnSuperPoint}
  \big(\hspace{1pt}
    \overline{\psi}
    \,\Gamma^{a_1 a_2}\,
    \psi
  \big)
  ,\,
  \big(\hspace{1pt}
    \overline{\psi}
    \,\Gamma^{a_1 \cdots a_5}\,
    \psi
  \big)
  \;\in\;
  \mathrm{CE}\big(
    \mathbb{R}^{0 \vert \mathbf{32}}
  \big)
  \,,
  \hspace{1cm}
  a_i \in \{0, \cdots, 10\}
  \,,
\end{equation}
since the spinor-valued 1-forms $\psi^\alpha$ are of bi-degree $(1,\mathrm{odd})$, hence mutually commuting \eqref{TheSignRule}, and since \eqref{Further2CocyclesOnSuperPoint} are the only {\it symmetric} $\mathrm{Spin}(1,10)$-invariant pairings \eqref{SymmetricSpinorPairings}.

\smallskip

Therefore the {\it maximal} $\mathrm{Spin}(1,10)$-invariant central extension of the super-point $\mathbb{R}^{0\,\vert\,\mathbf{32}}$ has further central generators $Z^{a_1 a_2}$, $Z^{a_1 \cdots a_5}$ (skew-symmetric in their indices), corresponding to \eqref{Further2CocyclesOnSuperPoint},
\begin{equation}
  \label{TheBasicMAlgebra}
  \mathfrak{M}
  \;\;\simeq\;\;
  \mathbb{R}
  \Big\langle
    \grayunderbrace{
      (Q_\alpha)_{\alpha=1}^{32}
    }{
      \mathrm{deg}\,=\, (0,\mathrm{odd})}
    \,,
    \grayunderbrace{
      (P_a)_{a=0}^{10}
    }{
      \mathrm{deg}\,=\,(0,\mathrm{evn})
    }
    \,,
    \grayunderbrace{
      (Z^{a_1 a_2} = Z^{[a_1 a_2]})_{a=0}^{10}
    }{
      \mathrm{deg}\,=\,(0,\mathrm{evn})
    }
    \,,
    \grayunderbrace{
      (Z^{a_1 \cdots a_5} = Z^{[a_1 \cdots a_5]})_{a=0}^{10}
    }{
      \mathrm{deg}\,=\,(0,\mathrm{evn})
    }
  \Big\rangle
\end{equation}

\vspace{0mm} 
\noindent with non-vanishing super-Lie bracket on generators now given by \footnote{
  The signs in  \eqref{TheFullyExtendedSusyBracket} are conventional --- for our choice of the first sign see ftn. \ref{PrefactorConvention} below, and for the second sign see ftn. \ref{TheSignInde}.
}
\begin{equation}
  \label{TheFullyExtendedSusyBracket}
  \begin{array}{c}
    [Q_\alpha,\,Q_\beta]
    \;=\;
    -
    \,
    2 \,\Gamma^a_{\alpha \beta}\, P_a
    \,
    \,+\,
    2 \,\Gamma^{a_1 a_2}_{\alpha \beta}\, Z_{a_1 a_2}
    \,
    \,-\,
    2 \,\Gamma^{a_1 \cdots a_5}_{\alpha \beta}\, Z_{a_1 \cdots a_5}
    \,.
  \end{array}
\end{equation}

\smallskip 
This fully brane-extended version of (the translational part of) the $D=11$, $\mathcal{N}=1$ supersymmetry algebra may be understood (\cite{dAGIT89}\cite[(13)]{Townsend95}\cite[(1)]{Townsend98}, cf. also  \cite{SS17-BPS}) as incorporating charges $Z^{a_1 a_2}$ of M2-branes and $Z^{a_1 \cdots a_5}$ of M5-branes, whence we shall call this the {\it basic M-algebra}, following \cite{Sezgin97}\cite{BDPV05}\cite[(3.1)]{Bandos17}. 
\footnote{
  \cite{Sezgin97} uses the term ``M-algebra'' for a large further extension of \eqref{TheFullyExtendedSusyBracket} which includes the original ``hidden M-algebra'' of \cite{DF82}; whereas other authors like \cite{BDPV05} say ``M-algebra'' for just \eqref{TheFullyExtendedSusyBracket}. Here we disambiguate this situation by speaking of the ``basic'' M-algebra and its ``hidden'' extension.
} 

Its CE-algebra is 
\begin{equation}
  \label{CEOfBasicMAlgebra}
  \mathrm{CE}\big(
    \mathfrak{M}
  \big)
  \;\;
  \;\simeq\;
  \mathbb{R}\Big[
    \grayunderbrace{
    (\psi^\alpha)_{\alpha=1}^{32}
    }{ 
      \mathrm{deg} = (1,\mathrm{odd})
    }
    ,\,
    \grayunderbrace{
    (e^a)_{a=0}^{10}
    }{
      \mathrm{deg} = (1,\mathrm{evn})
    }
    ,\,
    \grayunderbrace{
    (e_{a_1 a_2} = e_{[a_1 a_2]})_{a_i=0}^{10}
    }{
      \mathrm{deg} = (1,\mathrm{evn})
    }    
    \,,\,
    \grayunderbrace{
    (e_{a_1 \cdots a_5} 
      = 
    e_{[a_1 \cdots a_5]})_{a_i=0}^{10}
    }{
      \mathrm{deg} = (1,\mathrm{evn})
    }
  \Big]
  \,,
\end{equation}
with differential on generators given by
\footnote{
\label{TheSignInde}
We have a minus sign in the equation for $\mathrm{d}\, e_{a_1 a_2}$ in \eqref{DifferentialOnCEOfBasicMAlgebra} to match the sign convention in \cite[(6.2)]{DF82}\cite[(17)]{BDIPV04}, which is natural in view of \eqref{TheBispinorCEElement} below, and hence ultimately due to the relative sign in the formula \eqref{FierzDecomposition} for Fierz expansion.}

\vspace{-.4cm}
\begin{equation}
  \label{DifferentialOnCEOfBasicMAlgebra}
  \def\arraystretch{1.3}
  \begin{array}{lcl}
  \mathrm{d}\, 
  \psi 
  &=&
  0
  \\
  \mathrm{d}\,
  e^a 
  &=&
  +
  \big(\, 
    \overline{\psi}
    \,\Gamma^a\,
    \psi
  \big)
  \\
  \mathrm{d}
  \,
  e_{a_1 a_2}
  &=&
  -
  \big(\,
    \overline{\psi}
    \,\Gamma_{a_1 a_2}\,
    \psi
  \big)
  \\
  \mathrm{d}
  \,
  e_{a_1 \cdots a_5}
  &=&
  +
  \big(\,
    \overline{\psi}
    \,\Gamma_{a_1 \cdots a_5}
    \psi
  \big)
  \,.
  \end{array}
\end{equation}

Therefore the basic M-algebra \eqref{TheBasicMAlgebra} may be understood as a central extension of the super-point by the exceptional tangent space \eqref{ExceptionalTangentBundleInIntroduction}:
\begin{equation}
  \label{BasicMAlgebraAsCentralExtension    }
  \hspace{-1.5cm}
  \begin{tikzcd}[
    sep=15pt
  ]
  0
  \ar[r]
  &
  \mathbb{R}^{1,10}
  \oplus
  \wedge^2 (\mathbb{R}^{1,10})^\ast
  \oplus
  \wedge^5 (\mathbb{R}^{1,10})^\ast
  \ar[
    rr,
    hook
  ]
  &\phantom{--}&
  \mathfrak{M}
  \ar[rr, ->>]
  &\phantom{--}&
  \mathbb{R}^{0\,\vert\,\mathbf{32}}
  \ar[r]
  &
  0
  \end{tikzcd}
\end{equation}
In fact, this short exact sequence is itself an extension of the previous short exact sequence \eqref{SuperMinkowskiAlgebrasAsCentralExtension},
making the following commuting diagram of super-Lie algebras, where all horizontal and all vertical rows are exact:
\begin{equation}
  \label{ExactSequences}
  \begin{tikzcd}[
   column sep=12pt,
   row sep=6pt
  ]
  &
  0 \ar[d]
  &&
  0 \ar[d]
  &&
  0 \ar[d]
  \\
  0
  \ar[r]
  &
  \wedge^2 (\mathbb{R}^{1,10})^\ast
  \oplus
  \wedge^5 (\mathbb{R}^{1,10})^\ast
  \ar[
    rr, 
    equals,
    "{
      \scalebox{.7}{
        \color{darkblue}
        M-brane charges
      }
    }"{yshift=10pt}
  ]
  \ar[dd, hook]
  &&
  \wedge^2 (\mathbb{R}^{1,10})^\ast
  \oplus
  \wedge^5 (\mathbb{R}^{1,10})^\ast  
  \ar[dd, hook]
  \ar[rr, ->>]
  &&
  0
  \ar[r]
  \ar[dd, hook]
  &
  0
  \\
  \\
  0
  \ar[r]
  &
  \mathbb{R}^{1,10}
  \oplus
  \wedge^2 (\mathbb{R}^{1,10})^\ast
  \oplus
  \wedge^5 (\mathbb{R}^{1,10})^\ast
  \ar[
    rr,
    hook
  ]
  \ar[
    dd,
    ->>
  ]
  &\phantom{--}&
  \mathfrak{M}
  \ar[rr, ->>]
  \ar[dd, ->>]
  &\phantom{--}&
  \mathbb{R}^{0\,\vert\,\mathbf{32}}
  \ar[r]
  \ar[
    dd,
    equals
  ]
  &
  0
  \\
  \\
  0 \ar[r]
  &
  \mathbb{R}^{1,10}
  \ar[
    rr,
    hook
  ]
  \ar[d]
  &&
  \mathbb{R}^{1,10\,\vert\,\mathbf{32}}
  \ar[
    rr,
    ->>
  ]
  \ar[d]
  &&
  \mathbb{R}^{0\,\vert\, \mathbf{32}}
  \ar[r]
  \ar[d]
  &
  0
  \\
  &
  0 && 0 && 0
  \\[-11pt]
  & 
  \mathclap{
  \scalebox{.7}{
    \color{darkblue}
    \begin{tabular}{c}
      Generalized
      tangent space extension
      \\
      of Minkowski spacetime
    \end{tabular}
  }
  }
  &&
  \mathclap{
  \scalebox{.7}{
    \color{darkblue}
    \begin{tabular}{c}
      Basic M-algebra extension of
      \\
      super-Minkowski spacetime
    \end{tabular}
  }
  }
  &&
  \mathclap{
  \scalebox{.7}{
    \color{darkblue}
    \begin{tabular}{c}
      Odd tangent space 
    \end{tabular}
  }
  }
  \end{tikzcd}
\end{equation}
This gives a precise sense in which the basic M-algebra is a super-space analog of the local model for the generalized tangent space in M-theory.

\smallskip 
What remains to be seen is then that  $\mathfrak{k}_{1,10}$ acts, through its quotient $\mathfrak{sl}_{32}$, on the M-algebra. This was observed by \cite[\S 4]{West03}, following \cite[\S 5]{BaerwaldWest00}, using adapted Lie generators. Here we offer the  following streamlined argument, following \cite{BDIPV04}, which also shows that the full automorphism algebra is $\mathfrak{gl}_{32}$:

\medskip

\noindent
{\bf  $\mathrm{GL}(32)$-Automorphisms of the M-algebra.}\label{ManifestGL32}
  Unifying all the bosonic generators of \eqref{CEOfSuperMinkowksi}
  into a symmetric bispinorial form like this

  \vspace{-.4cm}
  \begin{equation}
    \label{TheBispinorCEElement}
    e^{\alpha \beta}
    \;\;
    :=
    \;\;
    \tfrac{1}{32}
    \big(
      e^a
      \,
      \Gamma
        _a 
        ^{\alpha \beta}
      +
      \tfrac{1}{2}
      e^{a_1 a_2}
      \,
      \Gamma
        _{a_1 a_2}
        ^{\alpha \beta}
      +
      \tfrac{1}{5!}
      e^{a_1 \cdots a_5}
      \,
      \Gamma
        _{a_1 \cdots a_5}
        ^{\alpha \beta}
    \big)
  \end{equation}
  the differential \eqref{DifferentialOnCEOfBasicMAlgebra}
  acquires equivalently the compact form
  \begin{equation}
    \label{ManifestEquivariantDifferentialOnBasicMAlgebra}
    \def\arraystretch{1.2}
    \begin{array}{lcl}
      \mathrm{d}
      \,
      \psi^\alpha
      &=&
      0
      \\
      \mathrm{d}
      \,
      e
        ^{\alpha \beta}
      &=&
      \psi^\alpha
      \,
      \psi^\beta
      \mathrlap{\,,}
    \end{array}
  \end{equation}
  which makes manifest that any $g\in \mathrm{GL}(32)$ acts via super-Lie algebra automorphisms of the M-algebra 
  \begin{equation}
    \label{ManifestGL32Equivariance}
    \begin{tikzcd}[row sep=-3pt,
      column sep=0pt
    ]
      \mathllap{
        g
        \;:\;
      }
      \mathrm{CE}\big(
        \mathfrak{M}
      \big)
      \ar[rr]
      &&
      \mathrm{CE}\big(
        \mathfrak{M}
      \big)
      \\
      \psi^\alpha
      &\longmapsto&
      g^{\alpha}_{\alpha'}
      \,
      \psi^{\alpha'}
      \\
      e^{\alpha \beta}
      &\longmapsto&
      g^{\alpha}_{\alpha'}
      \,
      g^{\beta}_{\beta'}
      \,
      e^{\alpha'\beta'}
      \mathrlap{\,.}
    \end{tikzcd}
  \end{equation}
Using our approach, we may succinctly show this as follows. 
 First, to see that the transformation \eqref{TheBispinorCEElement} is invertible,
  the trace-property
  \eqref{VanishingTraceOfCliffordElements}
  allows to recover:
  \begin{equation}
    \label{OriginalBosonicGeneratorsFromManifestlyGL32EquivariantBasis}
    \def\arraystretch{1.4}
    \begin{array}{lcl}
      e^a
      \;=\;
      \phantom{+}
      \Gamma
        ^a
        _{\alpha \beta}
      \,
      e^{\alpha \beta}
      \,,
      \hspace{.5cm}
      e^{a_1 a_2}
      \;=\;
      -
      \Gamma
        ^{a_1 a_2}
        _{\alpha \beta}
      \,
      e^{\alpha \beta}
      \,,
      \hspace{.5cm}
      e^{a_1 \cdots a_5}
      \;=\;
      \phantom{+}
      \Gamma
        ^{a_1 \cdots a_5}
        _{\alpha \beta}
      \,
      e^{\alpha \beta}.
    \end{array}
  \end{equation}
  The main point then is that the differential is as claimed, which follows by the Fierz expansion formula \eqref{FierzDecomposition}:
  $$
    \def\arraystretch{1.5}
    \begin{array}{ccll}
      \mathrm{d}
      \;
      e^{\alpha \beta}
     &=&
    \tfrac{1}{32}
    \Big(
      \Gamma
        _a 
        ^{\alpha\beta}
      \,
      \big(\hspace{1pt}
        \overline{\psi}
        \,\Gamma^a\,
        \psi
      \big)
      -
      \tfrac{1}{2}
      \Gamma
        _{a_1 a_2}
        ^{\alpha\beta}
      \big(\hspace{1pt}
        \overline{\psi}
        \,\Gamma^{a_1 a_2}
        \,
        \psi
      \big)
      +
      \tfrac{1}{5!}
      \Gamma
        _{a_1 \cdots a_5}
        ^{\alpha\beta}
      \big(\hspace{1pt}
        \overline{\psi}
        \,\Gamma^{a_1 \cdots a_5}\,
        \psi
      \big)
    \Big)
    &
    \proofstep{
     by
     \eqref{TheBispinorCEElement}
     \&
     \eqref{DifferentialOnCEOfBasicMAlgebra}
    }
    \\
    &=&
    \psi^\alpha
    \,
    \psi^{\beta}
    &
    \proofstep{
      by 
      \eqref{FierzDecomposition}.
    }
    \end{array}
  $$

\medskip

\noindent
{\bf Brane rotating symmetry.}\label{BraneRotatingSymmetry} 
On the original bosonic generators \eqref{CEOfBasicMAlgebra} -- the spacetime momentum $e^a$, the M2-brane charges $e^{a_1 a_2}$ and the M5-brane charges $e^{a_1 \cdots a_5}$ --- the 
 $\mathrm{GL}(32)$ symmetry of \eqref{ManifestEquivariantDifferentialOnBasicMAlgebra} 
 acts by mixing them all among each other, e.g.
 $$
   \def\arraystretch{1.4}
   \begin{array}{lcll}
     e^a
     &=&
     \Gamma^a_{\alpha \beta}
     e^{\alpha \beta}
     \;\overset{g}{\mapsto}\;
     \Gamma^a_{\alpha \beta}
     g^\alpha_{\alpha'}
     g^{\beta}_{\beta'}
     e^{\alpha' \beta'}
     &
     \proofstep{
       by  \eqref{OriginalBosonicGeneratorsFromManifestlyGL32EquivariantBasis}
       \eqref{ManifestGL32Equivariance}
     }
     \\
     &=&
     \big(
     \tfrac{1}{32}
     \Gamma^a_{\alpha \beta}
     g^\alpha_{\alpha'}
     g^{\beta}_{\beta'}
     \Gamma_b^{\alpha'\beta'}
     \big)
     e^b
     \,+\,
     \big(
     \tfrac{1}{64}
     \Gamma^a_{\alpha \beta}
     g^\alpha_{\alpha'}
     g^{\beta}_{\beta'}
     \Gamma_{b_1 b_2}^{\alpha'\beta'}
     \big)
     e^{b_1 b_2}
     \,+\,
     \big(
     \tfrac{1}{5!\cdot 32}
     \Gamma^a_{\alpha \beta}
     g^\alpha_{\alpha'}
     g^{\beta}_{\beta'}
     \Gamma_{b_1 \cdots b_5}
       ^{\alpha'\beta'}
     \big)
     e^{b_1 \cdots b_5}
     &
     \proofstep{
       by
       \eqref{TheBispinorCEElement},
     }
   \end{array}
 $$
 as befits an exceptional-geometric symmetry. For this reason, the authors \cite{BaerwaldWest00} speak of a ``brane rotating symmetry''.

\smallskip 
  By the discussion in \S\ref{ExceptionalTangentSpaces},
  this enhanced equivariance 
  \eqref{ManifestGL32Equivariance}
  of the M-algebra, which makes the basic super-Lie bracket a morphism of $\mathfrak{sl}_{32}$-representations
  $\mathbf{32} \otimes_{\mathrm{sym}} \mathbf{32} \;\simeq\; \mathbf{526}$, will have to be understood as the effective part of the corresponding $\mathfrak{k}_{1,10}$-action, according to \cite[p. 42]{BossardKleinschmidtSezgin19}.

\section{Conclusions}
\label{Vistas}

\noindent
{\bf The super-exceptional tangent space.} 
We have observed that the bosonic body of the M-algebra completes the pattern of the exceptional tangent spaces, traditionally discontinued at $n = 7$, all the way to $n = 11$, if one understands that the hidden U-duality symmetries which should act on local model spaces must be the {\it local} symmetries $\mathfrak{k}_n \subset \mathfrak{e}_n$, only --- and we concluded by highlighting that the resulting $\mathfrak{k}_{1,10}$-action lifts to the full M super-algebra in a way which completes familiar actions on exceptional $n$-tangent spaces from $n \leq 7$ to $n = 11$. 

\smallskip 
This lends support to the suggestion that the M-algebra serves as a combined super-exceptional tangent space for M-theory, as had previously been suggested, from different angles, in \cite{Vaula07}\cite{Bandos17}\cite[\S 4.5]{FSS20-HigherT}\cite{FSS20Exc}\cite{FSS21Exc}.

\smallskip 
While of course $\mathfrak{k}_{1,10}$ is vastly larger than its $\mathfrak{sl}_{32}$-quotient, \eqref{TowerOfExceptionalTangentBundles} suggests that the M-algebra may be understood as the local model space for a super-exceptional geometric formulation of 11D SuGra which retains the power of its (finite-dimensional) super-space formulation \cite{BH80}\cite{CF80}\cite{CDF91}\cite{GSS24-SuGra} while making manifest as much as possible of hidden U-duality symmetry, via the resulting ``super-exceptional Poincar{\'e} group'': 
\footnote{
  The super-Lie algebra of our ``super-exceptional Poincar{\'e} group''  has been referred to in \cite[\S 3.1]{MSCZ21} as ``an almost universal maximal supersymmetry algebra''.
}

\vspace{-.5cm}
\begin{center}
\hypertarget{TableKleinian}{}
\def\tabcolsep{2pt}
\begin{tabular}{ll}
\small
\hspace{-4mm}
\def\arraystretch{2}
\def\tabcolsep{2pt}
\begin{tabular}{|c|c|c|}
  \hline
  \bf
  \def\arraystretch{.9}
  \begin{tabular}{c}
    Kleinian local
    model space
    \\
    $S \,=\, G/H$
  \end{tabular}
  &
  \bf
  \def\arraystretch{.9}
  \begin{tabular}{c}
    Isometry 
    group
    \\
    $G$
  \end{tabular}  
  &
  \bf
  \def\arraystretch{.9}
  \begin{tabular}{c}
    Point
    group
    \\
    $H$
  \end{tabular}  
  \\
  \hline
  \hline
  \rowcolor{lightgray}
  \def\arraystretch{1}
  \begin{tabular}{c}
    Minkowski
    spacetime
    \\
    $\mathbb{R}^{1,10}$
  \end{tabular}
  &
  \def\arraystretch{1}
  \begin{tabular}{c}
    Poincar{\'e}
    group
    \\
    $\mathbb{R}^{1,10}
    \rtimes \mathrm{O}(1,10)$
  \end{tabular}
  &
  \def\arraystretch{1}
  \begin{tabular}{c}
    Lorentz
    group
    \\
    $\mathrm{O}(1,10)$
  \end{tabular}
  \\
  \rowcolor{white}
  \def\arraystretch{1}
  \begin{tabular}{c}
    Super-Minkowski spacetime
    \\
    $\mathbb{R}^{1,10\,\vert\,\mathbf{32}}$
  \end{tabular}  
  &
  \def\arraystretch{1}
  \begin{tabular}{c}
    Super-Poincar{\'e} group
    \\
    $\mathbb{R}^{1,10\,\vert\,\mathbf{32}} \rtimes \mathrm{Pin}^+(1,10)$
  \end{tabular}  
  &
  \def\arraystretch{1}
  \begin{tabular}{c}
    Pin-group
    \\
    $\mathrm{Pin}^+(1,10)$
  \end{tabular}  
  \\
  \rowcolor{lightgray}
  \def\arraystretch{1}
  \begin{tabular}{c}
    Super-exceptional spacetime
    \\
    (M-algebra)
    \\
    $\mathbb{R}^{
      1,10\,+\,517\,\vert\,\mathbf{32}
    }$
  \end{tabular}  
  &
  \def\arraystretch{1}
  \begin{tabular}{c}
    \color{darkgreen}
    \bf
    Super-exceptional
    \\
    \color{darkgreen}
    \bf
    Poincar{\'e} group
    \\
    $\mathbb{R}  
      ^{
       1,10\,+\,517
       \,\vert\,
       \mathbf{32}
    } 
    \rtimes 
    \mathrm{SL}(32)$
  \end{tabular}  
  &
  \def\arraystretch{.9}
  \begin{tabular}{c}
    Brane-rotating
    \\
    group
    \\
    $\mathrm{SL}(32)$
  \end{tabular}  
  \\
  \hline
\end{tabular}
&
$
  \begin{tikzcd}[
    row sep=10pt, 
    column sep=10pt
  ]
    &
    \overset{
    \mathclap{
      \raisebox{7pt}{
      \scalebox{.7}{
        \color{darkblue}
        \bf
        \def\arraystretch{.9}
        \begin{tabular}{c}
          Basic
          \\
          M-algebra
        \end{tabular}
      }
      }
    }    
    }{
    \overbrace{
      \mathbb{R}^{1,10\,+\,517\,\vert\, \mathbf{32}}
    }^{ 
      \mathfrak{M} 
    }
    }
    \ar[
      dl, 
      ->>,
      shorten=-1.2pt
    ]
    \ar[
      dr, 
      ->>,
      shorten=-1.2pt
    ]
    \ar[
      dd,
      ->>
      ]
    \\[-5pt]
    \mathbb{R}^{1,10\,\vert\,\mathbf{32}}
    \ar[
      dr, 
      ->>,
      shorten=-1.2,
      "{
      \scalebox{.7}{
        \color{darkblue}
        \bf
        \def\arraystretch{.9}
        \begin{tabular}{c}
          Super 
          \\
          tangent space
        \end{tabular}
      }      
      }"{swap, yshift=7pt}
    ]
    &&
    \mathbb{R}^{1,10\,+\,517}
    \ar[
      dl, 
      ->>,
      shorten=-1.2,
      "{
      \scalebox{.7}{
        \color{darkblue}
        \bf
        \def\arraystretch{.9}
        \begin{tabular}{c}
          Exceptional
          \\
          tangent space
        \end{tabular}
      }      
      }"{
        xshift=4pt,
        yshift=6pt
      }
    ]
    \\[-5pt]
    &
    \underset{
      \mathclap{
        \raisebox{-9pt}{
          \scalebox{.7}{
            \color{darkblue}
            \bf
            \def\arraystretch{.9}
            \begin{tabular}{c}
              Ordinary
              \\
              tangent space
            \end{tabular}
          }
        }
      }
    }{
      \mathbb{R}^{1,10}
    }
  \end{tikzcd}
$
\end{tabular}
\end{center}
\vspace{-.3cm}

But in consequence this means that for formulating 11D SuGra on ``super-exceptional spacetimes'',
the M-algebra (and its hidden extension) is to be understood not just as a super-Lie algebra but as a Kleinian model super-Lie {\it group}, carrying a left-invariant ``decomposed'' M-theory 3-form -- we discuss this in \cite{GSS24-HiddenGroup}.

\newpage

\appendix

\section*{Background}
\addtocounter{section}{1}

For ease of reference, we briefly state and cite some basic facts used in the main text.

\subsection{Spinors in 11D}
\label{SpinorsIn11d}

More details on the following may be found in
\cite[\S 2.5]{MiemiecSchnakenburg06}\cite[\S 2.2.1]{GSS24-SuGra}.

\smallskip

With Minkowski metric taken to be 
\begin{equation}
  \label{MinkowskiMetric}
  \big(\eta_{ab}\big)
    _{a,b = 0}
    ^{ d }
  \;\;
    =
  \;\;
  \big(\eta^{ab}\big)
    _{a,b = 0}
    ^{ d }
  \;\;
    :=
  \;\;
  \big(
    \mathrm{diag}
      (-1, +1, +1, \cdots, +1)
  \big)_{a,b = 0}^{d} \;.
\end{equation}
there exists an $\mathbb{R}$-linear representation $\mathbf{32}$ of $\mathrm{Pin}^+(1,10)$ with generators
\begin{equation}
  \label{The11dMajoranaRepresentation}
  \Gamma_a 
  \;:\;
  \mathbf{32}
  \xrightarrow{\;\;}
  \mathbf{32}
\end{equation}
equipped with a 
$\mathrm{Spin}(1,10)$-equivariant
skew-symmetric and non-degenerate
bilinear form
\begin{equation}
  \label{TheSpinorPairing}
  \big(\hspace{.8pt}
    \overline{(-)}
    (-)\,
  \big)
  \;:\;
  \mathbf{32}
  \otimes
  \mathbf{32}
  \xrightarrow{\quad}
  \mathbb{R}
\end{equation}
satisfying the following properties, where as 
usual we denote skew-symmetrized product of $k$ Clifford generators by
  \begin{equation}
    \label{CliffordBasisElements}
    \Gamma_{a_1 \cdots a_k}
    \;:=\;
    \tfrac{1}{k!}
    \underset{
      \sigma \in
      \mathrm{Sym}(k)
    }{\sum}
    \mathrm{sgn}(\sigma)
    \,
    \Gamma_{a_{\sigma(1)}}
    \cdot
    \Gamma_{a_{\sigma(2)}}
    \cdots
    \Gamma_{a_{\sigma(n)}}
    :
  \end{equation}

\begin{itemize}[leftmargin=.4cm]
  \item
  The Clifford generators square to the mostly plus Minkowski metric \eqref{MinkowskiMetric}
  \begin{equation}
    \label{CliffordDefiningRelation}
    \Gamma_a
    \Gamma_b
    +
    \Gamma_b
    \Gamma_a
    \;\;=\;\;
    +2 \, \eta_{a b}
    \,
    \mathrm{id}_{\mathbf{32}}
    \,.
  \end{equation}

  \item 
  The trace of all positive index Clifford basis elements vanishes:
  \begin{equation}
    \label{VanishingTraceOfCliffordElements}
    \mathrm{Tr}(
      \Gamma_{a_1 \cdots a_p}
    )
    \;=\;
    \left\{\!\!\!
    \def\arraystretch{1.3}
    \begin{array}{ccc}
      32 & \vert & p = 0
      \\
      0 & \vert & p > 0 
      \mathrlap{\,.}
    \end{array}
    \right.
    \,.
  \end{equation}

  \item
  The $\mathbb{R}$-vector space space of {\it symmetric} bilinear forms on $\mathbf{32}$
  has a linear basis given by the expectation values with respect to \eqref{TheSpinorPairing} of the 1-, 2-, and 5-index Clifford basis elements:
  \begin{equation}
    \label{SymmetricSpinorPairings}
    \mathrm{Hom}_{\mathbb{R}}
    \Big(
    (\mathbf{32}\otimes \mathbf{32})_{\mathrm{sym}}
    ,\,
    \mathbb{R}
    \Big)
    \;\;
    \simeq
    \;\;
    \Big\langle
    \big(
      (\overline{-})
      \Gamma_a
      (-)
    \big)
    \,,\;\;
    \big(
      (\overline{-})
      \Gamma_{a_1 a_2}
      (-)
    \big)
    \,,\;\;
    \big(
      (\overline{-})
      \Gamma_{a_1 \cdots a_5}
      (-)
    \big)
    \Big\rangle_{
      a_i = 0, 1, \cdots
      \,,
    }
  \end{equation}
  while a basis for the skew-symmetric bilinear forms is given by
  \begin{equation}
    \label{SkewSpinorPairings}
    \mathrm{Hom}_{\mathbb{R}}
    \Big(
    (\mathbf{32}\otimes \mathbf{32})_{\mathrm{skew}}
    ,\,
    \mathbb{R}
    \Big)
    \;\;
    \simeq
    \;\;
    \Big\langle
    \big(
      (\overline{-})
      (-)
    \big)
    \,,\;\;
    \big(
      (\overline{-})
      \Gamma_{a_1 a_2 a_3}
      (-)
    \big)
    \,,\;\;
    \big(
      (\overline{-})
      \Gamma_{a_1 \cdots a_4}
      (-)
    \big)
    \Big\rangle_{
      a_i = 0, 1, \cdots
      \,,
    }
  \end{equation}
\item which implies in particular the Fierz expansion
\begin{equation}
  \label{FierzDecomposition}
  \hspace{-3mm} 
  \big(\,
  \overline{\phi}_1
  \,
  \psi
  \big)
  \big(\,
  \overline{\psi}
  \,
  \phi_2
  \big)
  \;
  =
  \;
  \tfrac{1}{32}\Big(
    \big(\,
      \overline{\psi}
      \,\Gamma^a\,
      \psi
    \big)
    \big(\,
      \overline{\phi}_1
      \,\Gamma_a\,
      \phi_2
    \big)
    -
    \tfrac{1}{2}
    \big(\,
      \overline{\psi}
      \,\Gamma^{a_1 a_2}\,
      \psi
    \big)
    \big(\,
      \overline{\phi}_1
      \,\Gamma_{a_1 a_2}\,
      \phi_2
    \big)
    +
    \tfrac{1}{5!}
    \big(\,
      \overline{\psi}
      \,\Gamma^{a_1 \cdots a_5}\,
      \psi
    \big)
    \big(\,
      \overline{\phi}_1
      \,\Gamma_{a_1 \cdots a_5}\,
      \phi_2
    \big)
 \Big).
\end{equation}

\end{itemize}

\subsection{Super-Lie algebras}
\label{SuperLieAlgebras}

Our ground field is the real numbers $\mathbb{R}$, and all super-vector spaces are assumed to be finite-dimensional. Our notation follows \cite{FSS19-QHigher}, which gives more context.

\smallskip

\noindent
{\bf Sign rule.}
For homological super-algebra we consider bigrading in the direct product ring $\mathbb{Z} \times \ZTwo$ ---  where the first factor $\mathbb{Z}$ is the homological degree and the second $\ZTwo \simeq \{\mathrm{evn}, \mathrm{odd}\}$ the super-degree -- with sign rule
$$
  \mathrm{deg}_1 = 
  (n_1, \sigma_1),
  \;
  \mathrm{deg}_2
  =
  (n_2, \sigma_2)
    \,\in\,
  \mathbb{Z}\times \ZTwo
  \hspace{1cm}
    \yields
  \hspace{1cm}
  \mathrm{sgn}
  \big(
    \mathrm{deg}_1,
    \,
    \mathrm{deg}_2
  \big)
  \;:=\;
  (-1)^{n_1 \cdot n_2 + \sigma_1 \cdot \sigma_2}
  \,.
$$
(cf. e.g. \cite[p. 880]{BBLPT88}\cite[(II.2.109)]{CDF91} or \cite[\S 1]{DM99b}\cite[\S A.6]{DF99b}).

\smallskip

\noindent
{\bf Super-algebra.}
For $(v_i)_{i \in I}$ a set of generators with bi-degrees $(\mathrm{deg}_i)_{i \in I}$ we write:
\begin{itemize}[
  leftmargin=.7cm,
  topsep=2pt,
  itemsep=4pt
]
\item[\bf (i)]
$
  \mathbb{R}\big\langle
    (v_i)_{i \in I}
  \big\rangle
$
for the graded super-vector space spanned by these elements,
\item[\bf (ii)]
$
  \mathbb{R}\big[
    (v_i)_{i \in I}
  \big]
$
for the graded-commutative polymonial algebra generated by these elements, 

hence the tensor algebra on $\vert I\vert$ generators modulo the relation
\begin{equation}
  \label{TheSignRule}
  v_1 \cdot v_2
  \;=\;
  (-1)^{\mathrm{sgn}(
    \mathrm{deg}_1
    ,
    \mathrm{deg}_2
  )}
  \;
  v_2 \cdot v_1
  \,,
\end{equation}

i.e., the (graded, super) {\it symmetric algebra} on the above super-vector space:
$
  \mathbb{R}\big[ (v_i)_{i \in I} \big]
  :=
  \mathrm{Sym}
  \big(
    \mathbb{R}\big\langle 
      (v_i)_{i \in I} 
    \big\rangle
  \big).
$
\end{itemize}

\medskip

\noindent
{\bf Super-Lie algebra.}
Given a finite dimensional super-Lie algebra $\mathfrak{g} \,\simeq\, \mathfrak{g}_{\mathrm{evn}} \oplus \mathfrak{g}_{\mathrm{odd}}$,
the linear dual of the super-Lie bracket map
$$
  [\mbox{-},\mbox{-}]
  \;:\;
  \begin{tikzcd}
    \mathfrak{g}
    \vee 
    \mathfrak{g}
    \ar[r]
    &
    \mathfrak{g}
  \end{tikzcd}
$$
may be understood to map the first to the second exterior power of the underlying dual super-vector space, and as such it extends uniquely to a $\mathbb{Z}\!\times\!\ZTwo$-graded derivation $\mathrm{d}$ of degree=$(1,\mathrm{evn})$ on the exterior super-algebra (where the minus sign is just a convention)
$$
  \begin{tikzcd}[row sep=small]
    \wedge^1
    \mathfrak{g}^{\ast}
    \ar[
      rr,
      "{ 
        -[\mbox{-},\mbox{-}]^\ast
      }"
    ]
    \ar[
      d,
      hook
    ]
    &&
    \wedge^2 \mathfrak{g}^{\ast}
    \ar[
      d,
      hook
    ]
    \\
    \wedge^\bullet 
    \mathfrak{g}^\ast
    \ar[
      rr,
      "{ \mathrm{d} }"
    ]
    &&
    \wedge^\bullet 
    \mathfrak{g}^\ast    
    \,.
  \end{tikzcd}
$$
With this, the condition $d \circ d = 0$ is equivalently the super-Jacobi identity on $[\mbox{-},\mbox{-}]$, and
the resulting differential graded super-commutative algebra is know as the {\it Chevalley-Eilenberg algebra} of $\mathfrak{g}$:
$$
  \mathrm{CE}\big(
    \mathfrak{g}
    ,\,
    [\mbox{-},\mbox{-}]
  \big)
  \;\;
  :=
  \;\;
  \big(
    \wedge^\bullet \mathfrak{g}^\ast
    ,\,
    \mathrm{d}
  \big)
  \,.
$$
This construction is fully faithful
$$
  \begin{tikzcd}
    \mathrm{sLieAlg}_{\mathbb{R}}
    \ar[
      rr,
      hook,
      "{ \mathrm{CE} }"
    ]
    &&
    \mathrm{sDGCAlg}_{\mathbb{R}}
      ^{\mathrm{op}}
  \end{tikzcd}
$$
in that (1) for every super-vector space $V$ a choice of such differential $\mathrm{d}$ on $\wedge^\bullet V^\ast$ uniquely comes from a super-Lie bracket $[\mbox{-},\mbox{-}]$ on $V$ this way, and (2)
super-Lie homomorphisms $\phi : \mathfrak{g} \xrightarrow{\;} \mathfrak{g}'$ 
are in bijection with sDGC-algebra homomorphisms $\phi^\ast \,:\, \mathrm{CE}(\mathfrak{g}') \xrightarrow{} \mathrm{CE}(\mathfrak{g})$.
\begin{equation}
\label{RelationBetweenStructureConstants}\adjustbox{}{
\hspace{-.7cm}
\def\tabcolsep{4pt}
\begin{tabular}{p{8.15cm}l}
More concretely, given $(T_i)_{i =1}^n$ a linear basis for $\mathfrak{g}$ with corresponding structure constants $\big(f^k_{i j} \in \mathbb{R}\big)_{i,j,k = 1}^n$, then the Chevalley-Eilenberg algebra is the graded-commutative polynomial algebra
$
  \mathrm{CE}\big(
    \mathfrak{g}, 
    [\mbox{-},\mbox{-}]
  \big)
  \;\simeq\;
  \big(
    \mathbb{R}\big[
      t^1, \cdots, t^1
    \big]
    ,\,
    \mathrm{d}
  \big)
$
on generators of degree $(1,\sigma_i)$ with corresponding structure constants for its differential,
as shown on the right.
&
\;\;\;
\adjustbox{
  raise=-1.2cm
}{\small 
\def\arraystretch{1.7}
\begin{tabular}{|c||c|c|}
  \hline
  &
  \bf 
  \def\arraystretch{.95}
  \begin{tabular}{c}
    Super
    \\
    Lie algebra
  \end{tabular}
  &
  \bf 
  \def\arraystretch{.95}
  \begin{tabular}{c}
    Super
    \\
    dgc-algebra
  \end{tabular}
  \\
  \hline
  \hline
  Generators
  &
  $
    \big(
      \underbrace{
        T_i
      }_{ 
        \mathclap{
          \mathrm{deg} \,=\, (0,\sigma_i) 
        }
      }
    \big)_{i = 1}^n
  $
  &
  $
    \big(
      \underbrace{
        t^i
      }_{ 
        \mathclap{
          \mathrm{deg} \,=\, (1,\sigma_i) 
        }
      }
    \big)_{i = 1}^n
  $
  \\[-16pt]
  &&
  \\
  \hline
   \rowcolor{lightgray}
  Relations
  &
  $
    [T_i, T_j] 
    \,=\, 
    f^k_{i j}
    \,
    T_k
  $
  &
  $
    \mathrm{d}
    \,
    t^k
    \;=\;
    -
    \tfrac{1}{2}
    f^k_{i j}
    \, t^i t^j
  $
  \\
  \hline
\end{tabular}
}
\end{tabular}
}
\end{equation}

\end{document}